\documentclass[12pt]{iopart}
\usepackage{graphics}
\usepackage{graphicx}
\usepackage{epstopdf}
\usepackage{hyperref}
\begin{document}
\title[]{{\tt APFEL Web:} a Web-based application for the graphical
  visualization of parton distribution functions} \author{Stefano
  Carrazza$^{1,2}$\footnote{Corresponding author:
    \href{mailto:stefano.carrazza@mi.infn.it}{stefano.carrazza@mi.infn.it}},
  Alfio Ferrara$^3$, Daniele Palazzo$^2$\\ and Juan Rojo$^{4}$}

\address{$^1$Dipartimento di Fisica, Universit\`a degli Studi di
  Milano \& INFN, Sezione di Milano, Via Celoria 16, Milano, Italy}
\address{$^2$PH Department, TH Unit, CERN, CH-1211 Geneva 23,
  Switzerland} 
\address{$^3$Dipartimento di Informatica, Universit\`a
  degli Studi di Milano, Via Comelico 39, Milano, Italy}
\address{$^4$University of Oxford, 1 Keble Road, Oxford OX1 3NP,
  United Kingdom}
\ead{stefano.carrazza@mi.infn.it}
\date{Received: date / Revised version: date}
%
\begin{abstract}
  We present {\tt APFEL Web}, a Web-based application
  designed to provide a flexible user-friendly tool for the graphical
  visualization of parton distribution functions (PDFs). In this note
  we describe the technical design of the {\tt APFEL Web} application,
  motivating the choices and the framework used for the development of
  this project. We document the basic usage of {\tt APFEL Web} and
  show how it can be used to provide useful input for a variety of
  collider phenomenological studies. Finally we provide some examples
  showing the output generated by the application.
\end{abstract}
\pacs{89.20.Ff,12.38.-t}

\vspace{2pc}
\noindent{\it Keywords}: Parton distribution functions, LHC
phenomenology, graphical visualization
\section{Introduction}
\label{intro}

The requirements of precision physics at CERN's Large Hadron Collider
(LHC) have lead to the development of a variety of computer programs
that can be used to analyze LHC data and to provide predictions for
phenomenologically relevant processes.  The {\tt APFEL} library,
presented in Ref.~\cite{Bertone:2013vaa}, is an example of such a
program. {\tt APFEL} is a parton distribution function (PDF) evolution
library~\cite{Ball:2012wy,Ball:2013hta,Ball:2012cx,Ball:2014uwa,Harland-Lang:2014zoa}
which specialises in the solution of the DGLAP evolution equations
with QED corrections. The library also provides a native graphical
user interface ({\tt APFEL GUI}) with plotting tools for PDF
comparison, luminosities, deep-inelastic scattering (DIS) observables
and theoretical prediction computed through the {\tt APPLgrid}
interface~\cite{Carli:2010rw}. The {\tt APFEL GUI} reads PDF data
stored and interfaced by the {\tt LHAPDF}
library~\cite{Buckley:2014ana,Butterworth:2014efa} and plots are
produced with {\tt ROOT}~\cite{Brun:1997pa}.

After releasing the native version of the {\tt GUI} we have observed
that users had difficulties in maintaining the version of the PDF
grids available from the {\tt LHAPDF} library. Additionally the {\tt
  APFEL GUI} installation procedure required dependencies which are
constantly under development, e.g.~the {\tt Qt}
toolkit\footnote{\url{http://www.qt.io/}}, {\tt LHAPDF} and {\tt
  ROOT}. Therefore, in order to reduce the drawbacks of a native
application we have ported the code to an online centralized server
system that we called {\tt APFEL Web}. This service is designed with
the objective to provide a fast and complete interface to {\tt APFEL
  GUI} with an user-friendly Web-application interface. In this
respect, {\tt APFEL Web} provides also a timely replacement to the
{\tt HepData} online PDF
plotter\footnote{\url{http://hepdata.cedar.ac.uk/pdf/pdf3.html}}.

{\tt APFEL Web} is a Web-based application attached to a computer
cluster, available online at:
\begin{center}
{\bf \url{http://apfel.mi.infn.it/}~}
\end{center}
It contains PDF grids from {\tt LHAPDF5} and {\tt LHAPDF6}
libraries and it allows users to evolve PDFs using custom
configurations provided by the {\tt APFEL} library. Computational
results are presented in the format of plots which are produced by the
{\tt ROOT} framework.

This article is organized as follows. In Sect.~\ref{sec:design} we
document the application design and we explain the model scheme
developed for this project. In Sect.~\ref{sec:results} we discuss how
to use the Web-application and obtain results. Finally, in
Sect.~\ref{sec:conclusion} we present our conclusion and directions
for future work.

\section{Application design}
\label{sec:design}

The {\tt APFEL Web} project is divided into two parts: the server-side
and the cluster-side. The separation is a real requirement because the
service needs to interact with multiple users and computational jobs
at the same time. In the following we start from the description of
the Web framework developed for the server-side and then we show how
the combination is performed.

\subsection{The Web framework and interface}

\begin{figure}
  \begin{centering}
  \includegraphics[scale=0.7]{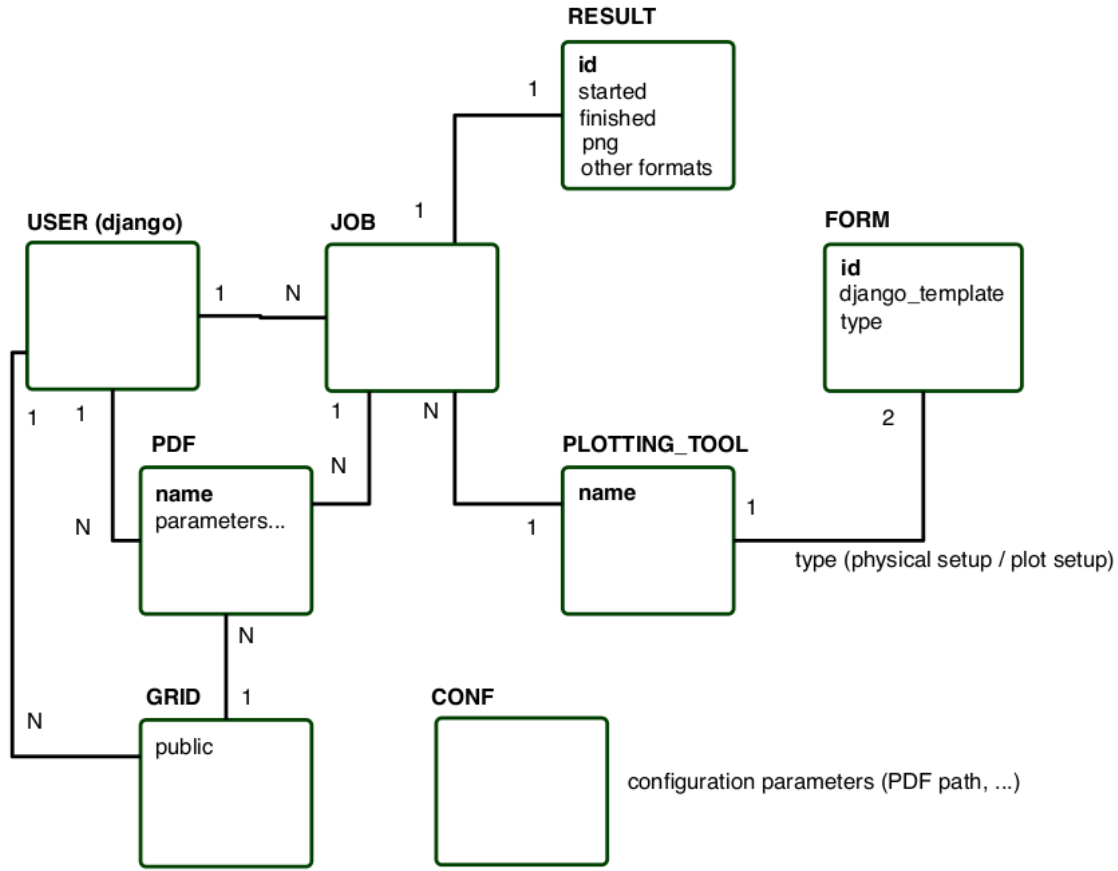}
  \par\end{centering}
\caption{A static design scheme of the {\tt APFEL Web} application
  model. The boxes represent a simplified view of the main components
  of this Web-application. Solid lines with 1/N labels represent the
  one/many relationships for each component of the application. Labels
  inside the boxes are examples of the database entry keys associated
  to the model.}
\label{fig:model}       
\end{figure}

For the development of the Web interface we have used the {\tt Django}
Web framework\footnote{\url{https://www.djangoproject.com/}}. {\tt
  Django} is a high-level {\tt Python} Web framework which provides a
high-performing solution for custom and flexible
Web-applications. Moreover the choice of {\tt Python} as programming
language instead of {\tt PHP} or {\tt Java}, is motivated by the need
of a simple interface to interact with the server system, by
simplifying the implementation of the communication between server and
cluster sides.

Following the {\tt Django} data model we have chosen to stored data in
a {\tt PostgreSQL}\footnote{\url{http://www.postgresql.org}} database
which should provide a good performance for our query requirements. We
use the authentication system provided by the {\tt Django} framework
in order to create a personal user Web-space, so users can save
privately personal configurations and start long jobs without need to
be connected over the whole calculation time.

In Figure~\ref{fig:model} we show a schematic view of the
Web-application model used in {\tt APFEL Web}. Starting from the
top-left element, users have access to {\tt PDF} objects which store
in the database the information about the PDF: e.g. the set name, the
PDF uncertainty treatment and the library for the treatment of PDF
evolution. Users have the option to choose PDF sets from the {\tt
  LHAPDF} library or, if preferred, upload their own private grid
using the {\tt LHAPDF5 LHgrid} and {\tt LHAPDF6} formats. Users are
able to run jobs after setting up the PDF grid objects: there are seven
job types which are classified in the image as plotting tools and will
be described in detail in Sect.~\ref{sec:results}. For each plotting
tool there are customized input Web-forms, implemented with the {\tt
  Django} {\tt models} framework, which collect information and store
it in the database before the job submission.  When a job finalizes,
it stores images and {\tt ROOT} files to the server disk, which are
then downloaded by the user. General configuration information such as
the path of the PDF grids and libraries are stored directly into the
{\tt Django} settings. 

Concerning Web-security in {\tt APFEL Web}, the user's account and its
information are protected by the {\tt Django Middleware}
framework. Undesirable users, such as spambots, are filtered by a
security question during the registration form. Finally, all users
have a limited disk quota which disable job submissions when exceeded.

\subsection{Computation engine and server deployment}

\begin{figure}
  \begin{centering}
  \includegraphics[scale=0.50]{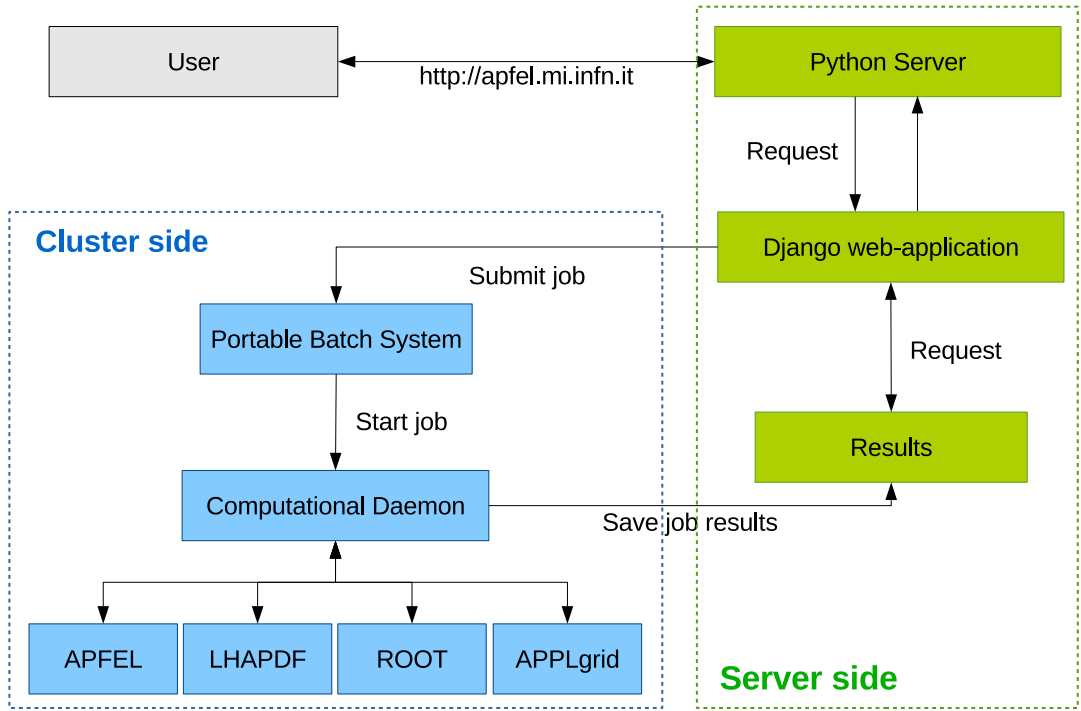}
  \par\end{centering}
\caption{Deployment layout of {\tt APFEL Web}.}
\label{fig:system}       
\end{figure}

In parallel to the Web development, the most important component of
{\tt APFEL Web} is the computational engine that we called {\tt
  apfeldaemon}. The program is a generalization of the open source
{\tt APFEL GUI} code in {\tt C++} with the inclusion of the database
I/O procedures. The job configuration and the PDF grids are read from
the database, and the computation is performed upon request by the
user. In order to solve the problem correlated with the usage of two
different interfaces to PDF grids, i.e. {\tt LHAPDF5} and {\tt
  LHAPDF6}, the {\tt apfeldaemon} is composed by two binaries which
are linked to the respective libraries: the Web-application checks the
PDF grid version and it starts the computation procedure with the
correct daemon.

In Figure~\ref{fig:system} we show the scheme of the Web-application
structure. Users from Web browsers send requests to a {\tt Python}
server which in our case is implemented by {\tt
  gunicorn} and {\tt nginx}\footnote{\url{http://gunicorn.org} and \url{http://nginx.org/}}. The {\tt Python} server
performs the request using the {\tt Django} framework, at this level
PDF objects and jobs are prepared and saved in the database,
additionally eventual job results are collected in a dedicated
view. From the computational point of view the layout is very simple
and clearly illustrated by the left side of Fig.~\ref{fig:system}. We
have set up a Portable Batch System (PBS)\footnote{An example of PBS
  open source implementation: \url{http://www.adaptivecomputing.com/}}
for the multi-core server which receives job submissions and is able
to automatically handle the job queue, avoiding the unpleasant
situation of server overloading. Jobs are submitted by the {\tt
  Django} application which passes the job identification number, this
value is read by the {\tt apfeldaemon} and it performs a query at the
corresponding database entry, then it collects the relevant
information to start the correct job. When the job finalizes the {\tt
  apfeldaemon} modifies the job status in the database, so the Web
interface notifies the user of the job status.

The {\tt apfeldaemon} program was designed and compiled with
performance as priority, in fact there are relevant computational
speed improvements when comparing to the previous {\tt APFEL GUI}
program almost due to the clear separation between the {\tt GUI} and
the calculation engine. In order to provide to the reader an idea of
the typical processing time per job, we estimate that one job requires
two seconds to process a single PDF set when producing a PDF
comparison plots, meanwhile for luminosity and observables jobs, the
system takes up to one minute per PDF set when including the
uncertainty treatment.

\section{Results}
\label{sec:results}

\begin{figure}
  \begin{centering}
    \includegraphics[scale=0.4]{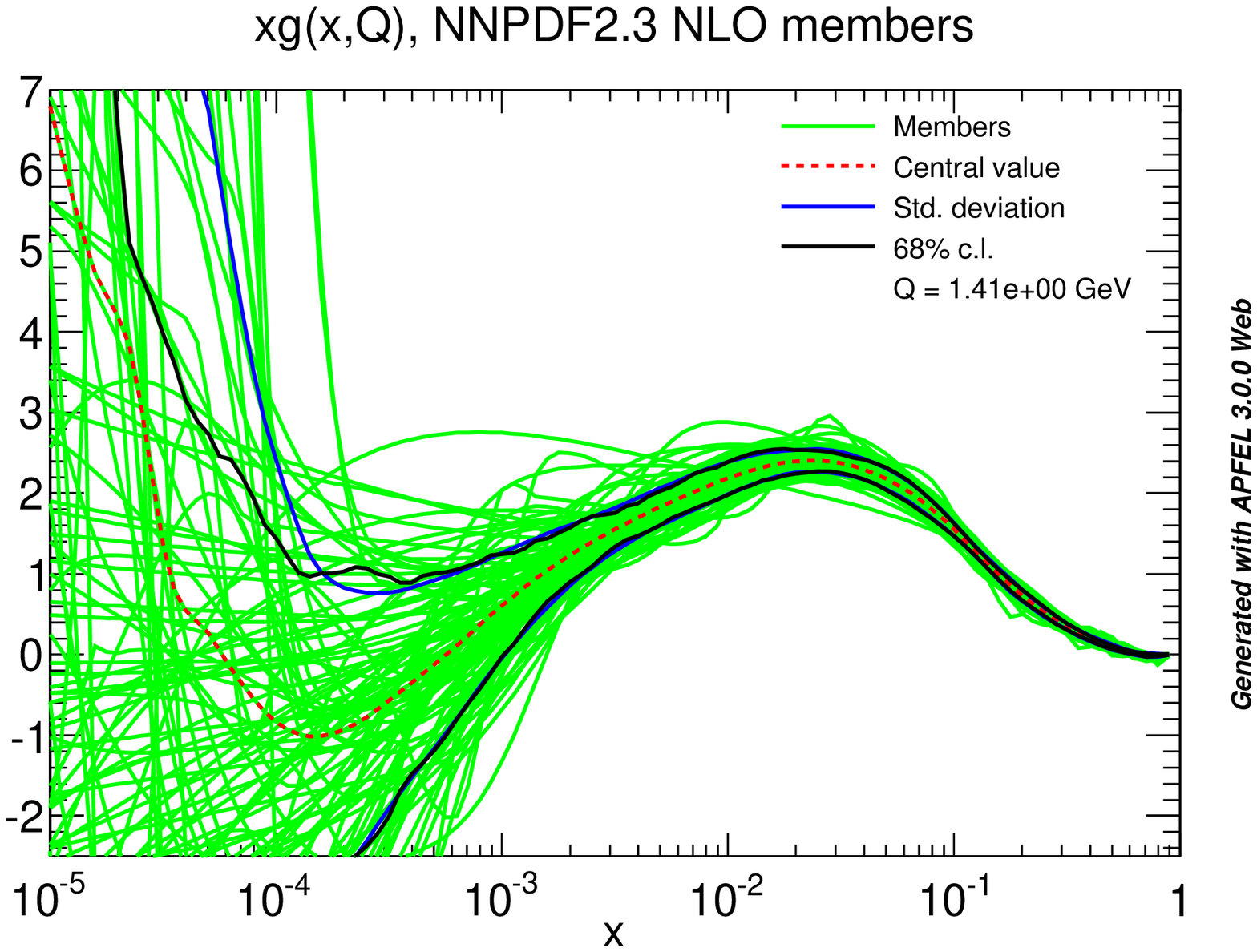}\includegraphics[scale=0.4]{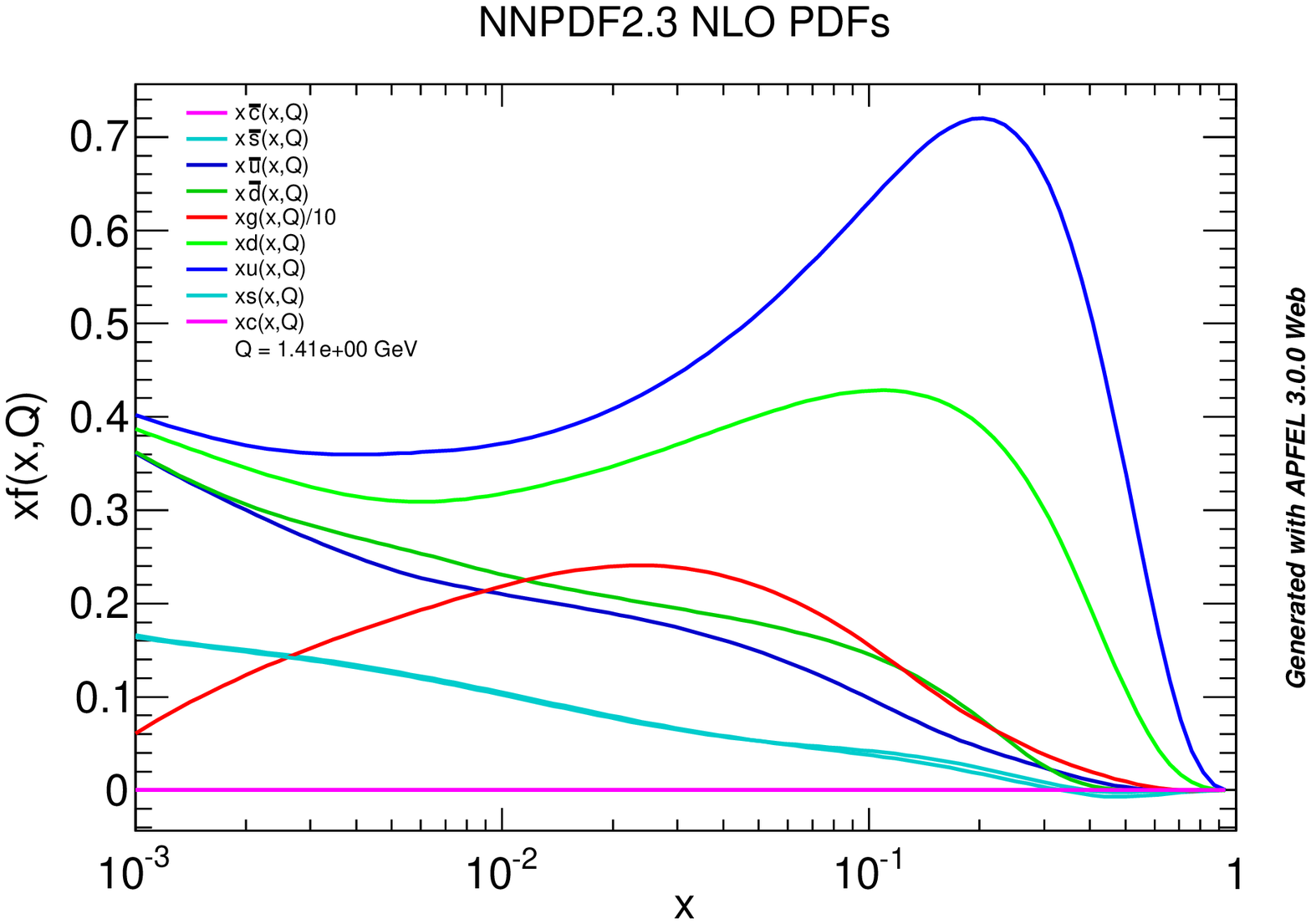}
    \includegraphics[scale=0.4]{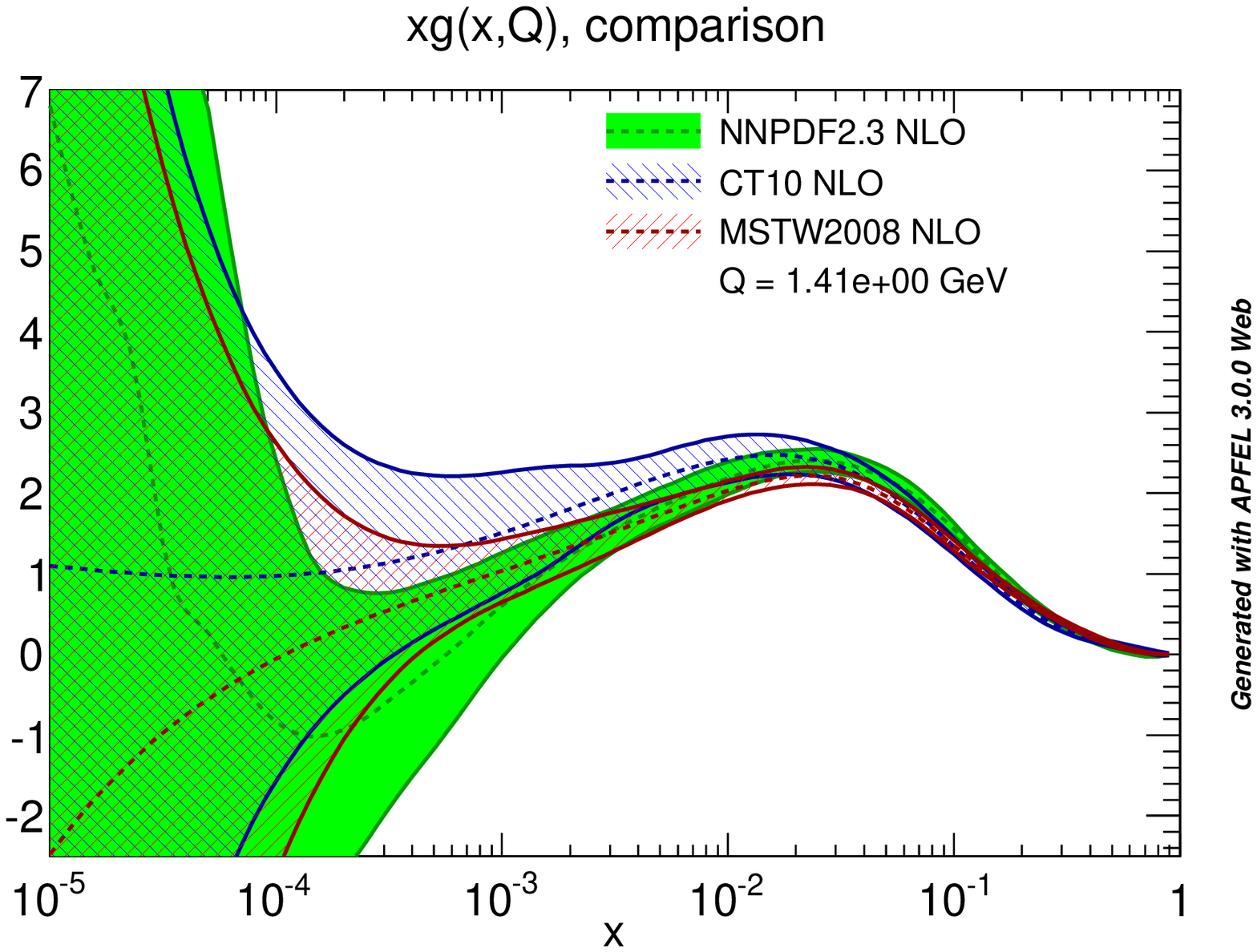}\includegraphics[scale=0.4]{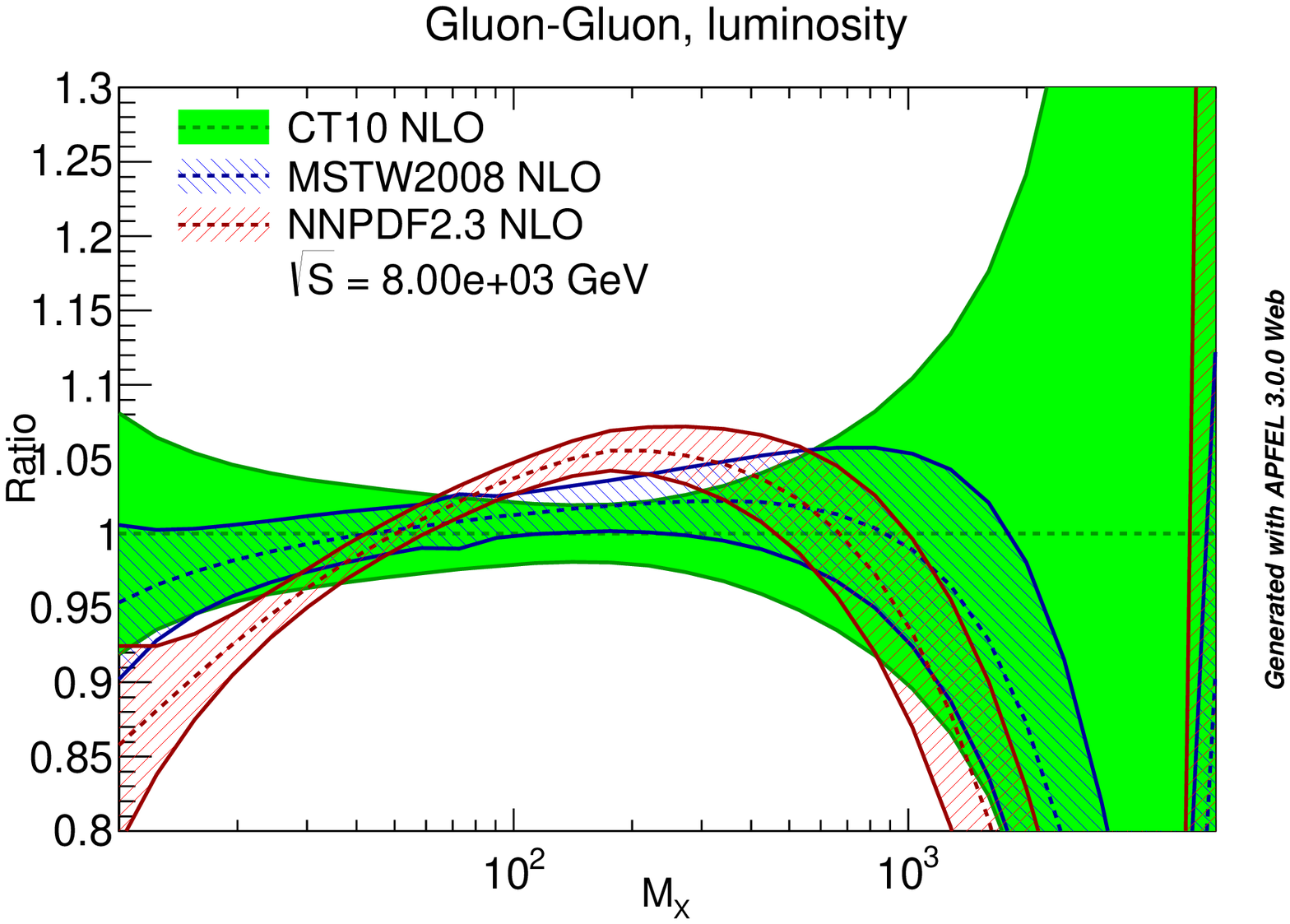}
    \par\end{centering}
  \caption{Examples of output generated with {\tt APFEL Web}. Plots
    are presented in the following order, clockwise from top-left:
    member plot, all flavors plot, PDF comparison and $gg$-channel
    luminosity.}
\label{fig:results}       
\end{figure}

While the use of the Web-interface should be self-explanatory, here we
describe and show examples of job results that a user is able to
obtain from {\tt APFEL Web}.

The first step consists in the creation of custom ``PDF objects'' in
the user's workspace. The following points explain how to create such
objects:
\begin{enumerate}

\item select a PDF grid from the {\tt LHAPDF5} and/or {\tt LHAPDF6}
  libraries and determine the treatment of the PDF uncertainty among:
  no error, Monte Carlo approach, Hessian eigenvectors (68 and 90\%
  c.l.)  and symmetric eigenvectors. When selecting a PDF set the
  system proposes automatically an uncertainty type based on the PDF
  collaboration name.

\item import a new {\tt LHAPDF} grid file, with the only requirement
  that it is provided either in the {\tt LHAPDF5 LHgrid} or in the
  {\tt LHAPDF6} format. The main target for this feature are the
  members of the PDF collaborations which can perform comparisons with
  preliminary sets of PDFs before the publication in {\tt LHAPDF}.

\item set the evolution library by choosing between the {\tt LHAPDF}
  interpolation routines or the {\tt APFEL} custom evolution.

\end{enumerate}

\begin{figure}
  \begin{centering}
    \includegraphics[scale=0.4]{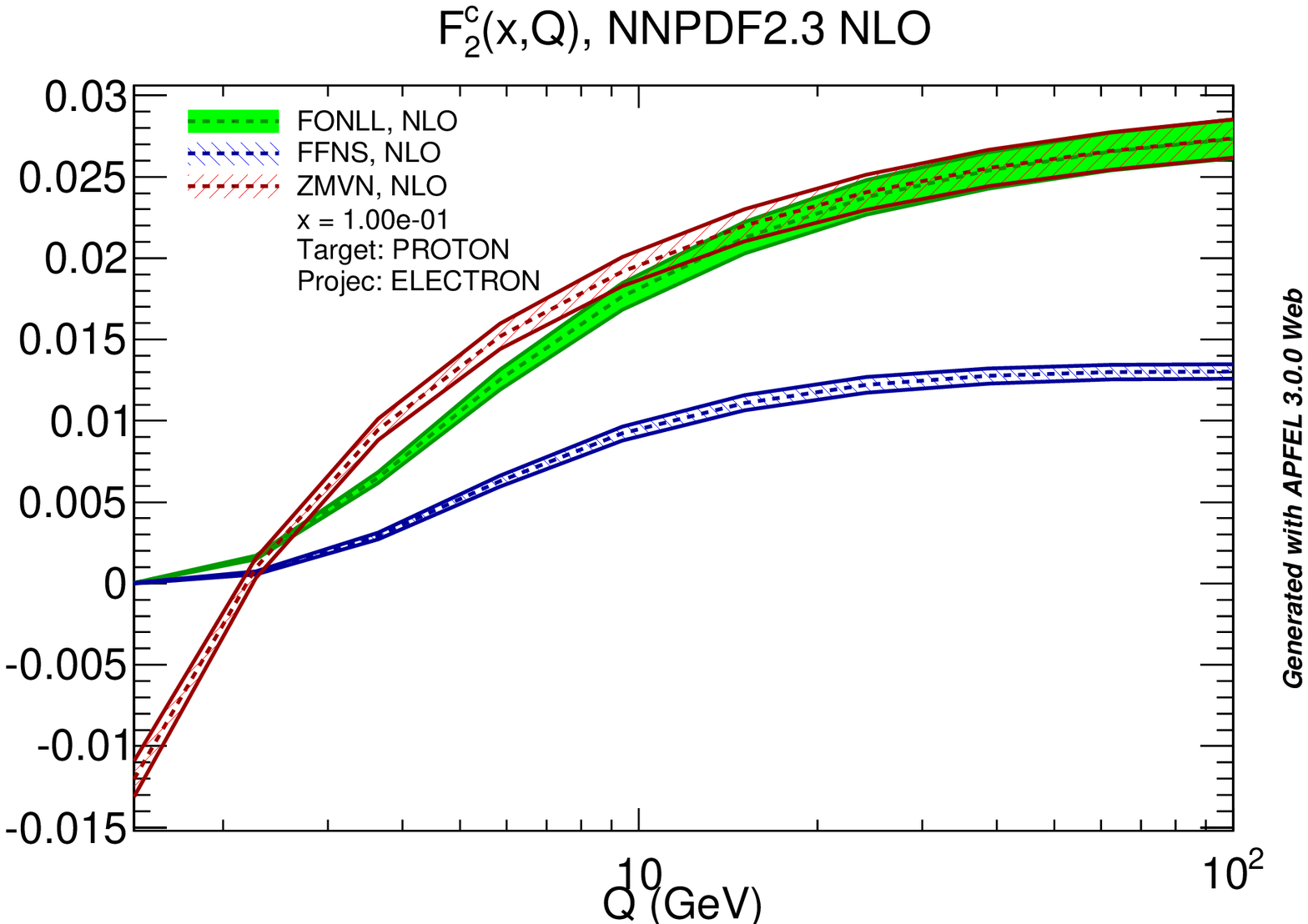}\includegraphics[scale=0.4]{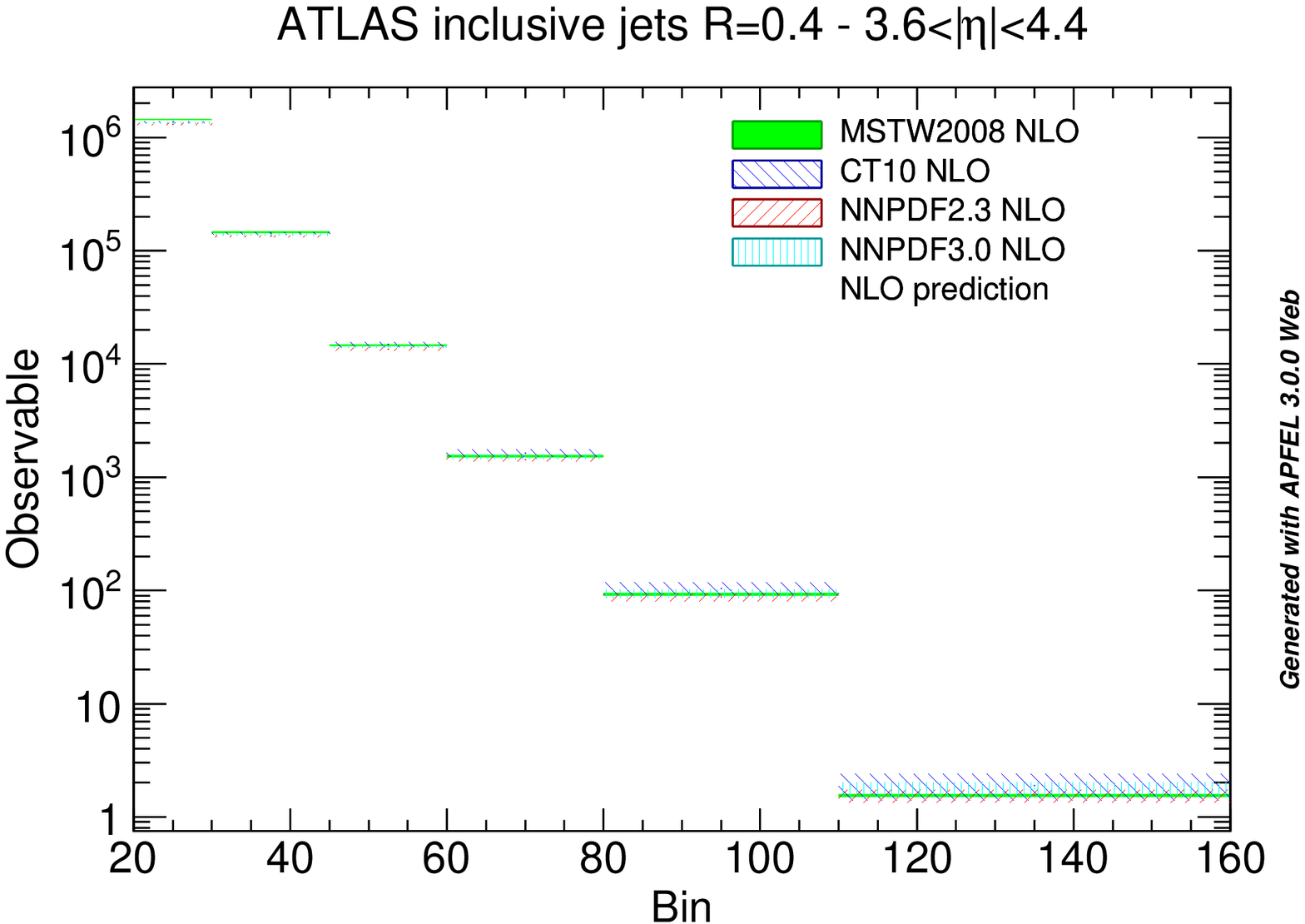}
  \par\end{centering}
\caption{On the left, an example of DIS observable computed by {\tt
    APFEL Web}: $F^{c}_{2}(x,Q)$. On the right, another example about
  the {\tt APPLgrid} observables tool used for the computation of
  predictions for ATLAS 2010 inclusive jets $R=0.4$ at $\sqrt{s}=7$
  TeV~\cite{Aad:2011fc}.}
\label{fig:results2}       
\end{figure}

We provide the following computational functions, which are
illustrated in Figures~\ref{fig:results} and~\ref{fig:results2}:
\begin{itemize}

\item ``{\tt Plot PDF Members}'': it plots for projections in $x$ all
  the members of a PDF set for a single parton flavor at a given
  energy scale $Q$. See the top-left image in Fig.~\ref{fig:results}
  where we show the replicas of {\tt NNPDF2.3 NLO}~\cite{Ball:2012cx}
  together with its central value and Monte Carlo uncertainty band,
  these last features are options which can be disabled by the
  user. This plotting tool accepts only a single PDF set at each time
  in order to avoid too many information in a single plot. We provide
  the possibility to choose between the usual parton flavors,
  i.e. $b,t,c,s,d,u,g,\gamma,q_i^{\pm}=q_i\pm\bar{q_i}$ with $q_i=u,d,s,c,b,t$, and the combination of them:
  ($\Sigma,V,V_{3},V_{15},V_{24},V_{35},T_{3},T_{15},T_{24},T_{35},\Delta_{s}$)~\cite{Ball:2011uy},
  the so called evolution basis.
  
\item ``{\tt Plot all PDF flavors}'': each PDF flavor is plotted
  together in the same canvas at a fixed energy scale. We also provide
  the possibility to scale PDF flavors by a predetermined numeric
  factor in order to produce plots similar to the
  PDG~\cite{Agashe:2014kda}. An example of PDF flavor plot is
  presented in the top-right of Fig.~\ref{fig:results} where the gluon
  PDF is scaled by a factor 10.
 
\item ``{\tt Plot Multiple PDFs (x)}'': this tool compares the same
  flavor of multiple PDF sets and the respective uncertainties at a
  given energy scale for projections in $x$. We provide the
  possibility to compute the absolute value or the just the ratio
  respect to a reference PDF set. The bottom-left image of
  Fig.~\ref{fig:results} shows the comparison between {\tt NNPDF2.3
    NLO}~\cite{Ball:2012cx}, {\tt CT10 NLO}~\cite{Lai:2010vv} and {\tt
    MSTW2008 NLO}~\cite{Martin:2009iq} sets at $Q=1$ GeV.

\item ``{\tt Plot Multiple PDFs (Q)}'': this tool compares the same
  flavor of multiple PDF sets and the respective uncertainties at a
  fixed $x$-value as a function of the energy scale $Q$.

\item ``{\tt Compute Luminosity}'': it performs the computation of
  parton luminosities~\cite{Ball:2010de} normalized to a reference PDF
  set at a given center of mass energy. There are several channels
  available: $gg$, $q\bar{q}$, $qg$, $cg$, $bg$, $qq$, $c\bar{c}$,
  $b\bar{b}$, $\gamma \gamma$, $\gamma g$, etc. In the bottom-right plot of
  Fig.~\ref{fig:results} we show an example of $gg$-luminosity at
  $\sqrt{s}=8$ TeV using the PDF sets presented above with {\tt CT10
    NLO} as reference PDF set.

\item ``{\tt Compute DIS(x)/DIS(Q)}'': it computes DIS observables as
  functions of $x$ or $Q$ for different heavy quark schemes and
  perturbative orders, including the Fixed Flavor Number scheme
  (FFNS), the Zero Mass Variable Number scheme (ZMVN), and the FONLL
  scheme~\cite{Forte:2010ta} where the choice of a NLO prediction
  implies using the FONLL-A scheme, while choosing NNLO leads to using
  the FONLL-C scheme. A detailed explanation of all possible
  configurations is presented in Sect.~4.3 of
  Ref.~\cite{Bertone:2013vaa}. An example of such tool is presented in
  the left plot of Fig.~\ref{fig:results2}.

\item ``{\tt APPLgrid observables}'': this tool provides a simple a
  fast interface to theoretical predictions through the {\tt APPLgrid}
  library~\cite{Carli:2010rw}. The system already provides several
  grids that are available from the {\tt APPLgrid}
  website\footnote{\url{http://applgrid.hepforge.org/}} but also from
  the NNPDF collaboration~\cite{Ball:2014uwa} and {\tt
    aMCfast}~\cite{Bertone:2014zva}.  This function allows users to
  compute the central value and the respective uncertainties for
  multiple PDF sets. On the right plot of Fig.~\ref{fig:results2} we
  show the output of this tool for the predictions of ATLAS 2010
  inclusive jets $R=0.4$ at $\sqrt{s}=7$ TeV~\cite{Aad:2011fc}.

\end{itemize}

For all the tools presented above, the Web interface provides options
for customizing the graphics, like setting the plot title, axis
ranges, axis titles and curve colors. {\tt APFEL Web} also provides
the possibility to save plots and the associated data in multiple
formats, including: PNG, EPS, PDF,~.C ({\tt ROOT}) and~.root ({\tt ROOT}).

Finally, it is important to highlight that the results produced by
{\tt APFEL Web} for PDF comparison and parton luminosities from
different PDF sets have been verified against the PDF benchmarking
exercise of Ref.~\cite{Ball:2012wy}.

\section{Summary and outlook}
\label{sec:conclusion}

{\tt APFEL Web} is a new Web-based application that provides a
user-friendly graphical user interface for the visualization of PDFs
with a wide range of formats: absolute plots, ratio plots, compare
PDFs from different groups, compare error PDF from a single set, plot
all PDF flavor combinations at the same time, compute parton
luminosities and finally compute also DIS structure functions and {\tt
  APPLgrid} observables. All these functionalities are accessed via a
centralized Web server.

{\tt APFEL Web} is available online for PC and mobile devices at:
\begin{center}
{\bf \url{http://apfel.mi.infn.it/}~}
\end{center}

The current framework provides a stable starting point to an future
expansion of the functions already implemented in {\tt APFEL Web}, the
design of the PDF objects is flexible enough to perform convolution of
PDFs in multiple external codes, suggesting the possibility to include
in future releases other software tools from HEP, enlarging the
functions of this Web-application.  

The {\tt APFEL Web} application was released on October 7, 2014. Four
months after the release we already have registered 76 users from 10
countries, with an average of 200 visits each month. Currently, the
server has successfully completed more than 900 jobs. Thanks to its
flexibility and user-friendliness, we believe that in the coming
months and years {\tt APFEL Web} has the potential to become a widely
used tool in the LHC community.

\subsection*{Acknowledgments}

We thank Stefano Forte and Valerio Bertone for intensive discussions
and feedback during the construction of this application. SC
acknowledges support by an Italian PRIN2010 grant and by an European
Investment Bank EIBURS grant.

\section*{References}
%
\bibliographystyle{jphysg}
\bibliography{apfelweb}

\end{document}